\documentclass[10pt]{article}
\usepackage{graphicx}
\usepackage{longtable}
\usepackage{times}
\usepackage{setspace}
\usepackage{dcolumn}
\usepackage{bm}
\usepackage{rotating}
\usepackage{hyperref}
\usepackage{epstopdf}
\usepackage{bm}
\usepackage{rotating}
\begin{document}
\begin{center}
    \textbf{The Variation of Photon Speed with Photon Frequency in Quantum Gravity}
\end{center}

\begin{center}
    $ Anuj \ Kumar \ Dubey^{*}$,  $ A. \ K. \ Sen$ and $ Sonarekha \ Nath$
\end{center}
\begin{center}
    \textit{Department of Physics, Assam University, Silchar-788011, Assam, India}
\end{center}
\begin{center}
*Corresponding Author, E-mail:danuj67@gmail.com
\end{center}
\begin{center}
(Accepted for publication in Indian Journal of Physics)
\end{center}
\begin{center}
Date: 10 March, 2018
\end{center}
\begin{abstract}
In the present work, an expression for Planck Mass or Planck Energy  is derived by equating the Compton wavelength with the gravitational radius of the Kerr rotating body.
Using the modified photon energy-momentum dispersion relation, the variation of the photon propagation speed with photon frequency is  derived.
It  is found that, the photon propagation speed, depends on the frequency of the photon, the rotation parameter of the Kerr rotating body and also on
the polarization state of the photon. Quantum gravity effect could be seen from the derived results for the photon propagation speed.
\end{abstract}

\section{Introduction}\label{section1}
In past it was  discussed by Rovelli  and Vidoto \cite{Rovelli-Vidoto, Vidoto2014}, that one can obtain the minimum size, where a
quantum particle can be localized without being hidden by its own horizon.

Accordingly the minimum size length (L), can be given as \cite{Rovelli-Vidoto, Vidoto2014}:
\begin{equation}\label{2}
    L=\frac{MG}{c^{2}}=\frac{EG}{c^{4}}=\frac{pG}{c^{3}}=\frac{\hbar G}{L c^{3}}
\end{equation}
Here G, $\hbar$ and c are the universal gravitational constant, planck's constant (over 2$\pi$) and speed of light respectively. M, p and E are mass, momentum
and energy respectively.

Now one can find that, it is not possible to localize  a quantum particle with a precision better than the  Planck length ($L_{Planck}$). The Planck length ($L_{Planck}$) can be expressed as:
\begin{equation}\label{3}
    L_{Planck}=\sqrt{\frac{\hbar G}{c^{3}}} \sim 1.6 \times 10^{-35} \ m
\end{equation}
 Similarly, it is also not possible to make measurement of time smaller than the Planck time ($t_{Planck}$) given as:
\begin{equation}\label{4}
t_{Planck}=\frac{L_{Planck}}{c}=(\frac{\hbar G}{c^{5}})^{\frac{1}{2}} \sim 5.4 \times 10^{-44} \ s
\end{equation}
The Planck mass ($M_{Planck}$) is given as:
\begin{equation}\label{5}
M_{Planck}=\frac{\hbar}{c \ L_{Planck}}=(\frac{\hbar c}{G})^{\frac{1}{2}} \sim 2.2\times10^{-8} \ kg
\end{equation}
 The Planck energy ($E_{Planck}$) is given as:
\begin{equation}\label{6}
E_{Planck}=M_{Planck} \ c^{2}=(\frac{\hbar c^{5}}{G})^{\frac{1}{2}} \sim 1.2\times10^{28} \ ev
\end{equation}

Einstein's Special Theory of Relativity is Lorentz invariant. Thus Special Theory of Relativity postulates that all observers measure
exactly the same speed of light in vacuum, independent of photon-energy.  Special Theory of Relativity does not assume any
fundamental length-scale associated with the space-time dynamics, while there is a fundamental scale the \textit{Planck scale},
at which quantum effects are expected to strongly affect the nature of space-time.  So there is a possibility that Lorentz
invariance might break near the Planck scale.  Due to this reason there is a great possibility for the variation of
photon speed with energy \cite{Abdo2009,Jacholkowska2015}.

From various approaches of quantum gravity, there are indications that the Lorentz invaraince is only an effective invariance which
holds in the low energy limit (say, for low-energy infrared photons) of quantum gravitational processes and this may be violated
in the high energy limit (say for Ultraviolet, X-ray, $\gamma$-ray photons) \cite{Carmona2003,Iengo2009,Carmona2006,Oriti2009}.
In the most popular approaches of quantum gravity, such as string theory \cite{Ellis2000grg,Ellis2000prd} and loop quantum gravity
\cite{Gambini-Pullin1999,Alfaro2000,Morales2002}, more general arguments are made that lead to the violations of Lorentz Invariance.
The fundamental symmetries, like Lorentz invariance or CPT, could be broken as discussed in the literature
\cite{Kostelecky1989,Tawfik,Kostelecky1991,Kostelecky1995,Kostelecky1996}.

The modification to the standard energy-momentum relation does occur in the various discrete space-time  models \cite{Hooft1996}
based on string field theory \cite{Kostelecky-Samuel1989}, space-time foam \cite{Amelino-Camelia1998}, the spin-network,
in loop quantum gravity \cite{Gambini-Pullin1999}, non-commutative geometry \cite{Carroll-Harvey2001}, and Horava-Lifshitz
gravity \cite{Horava2009}.

Magueijo and Smolin \cite{Magueijo-SmolinPRL2002,Magueijo-SmolinPRD2003} had given a modification of Special Theory of Relativity called
Double Special Relativity. In this theory of Double Special Relativity, in addition to the velocity of light being the maximum
velocity attainable, there is also a maximum energy scale the Planck energy $(E_{Planck})$, and it is not possible for a
particle to attain energies beyond this energy. The Double Special Relativity has been generalized to curved space-time, and
this Doubly General Theory of Relativity is called gravity's rainbow \cite{Magueijo-SmolinCQG2004}. In gravity's rainbow theory,
the geometry of space-time depends on the energy of the test particle. In order to construct this theory, the modified
energy-momentum dispersion relation in curved space-time has been discussed in details in the work of
Ali and Faizal \cite{Hendi-FaizalPRD2015, Ali-FaizalJHEP2014,Ali-FaizalNPB2015,Ali-FaizalPLB2015, Ashour-FaizalEPJ2016,
Ali-FaizalIJMMP2015,Ali-FaizalEPL2015} .

The modified energy-momentum dispersion relation can be written as \cite{Hendi-FaizalPRD2015, Ali-FaizalJHEP2014,
Ali-FaizalNPB2015,Ali-FaizalPLB2015,Ashour-FaizalEPJ2016,Ali-FaizalIJMMP2015,Ali-FaizalEPL2015} :
\begin{equation}\label{x}
    E^{2}f^{2} (\frac{E}{E_{Planck}})-p^{2} c^{2} g^{2} (\frac{E}{E_{Planck}}) = m^{2} c^{4}
\end{equation}
where $E_{Planck}$ is the Planck energy.

The functions f($\frac{E}{E_{Planck}}$) and g($\frac{E}{E_{Planck}}$) are called the rainbow functions, and they are
required to satisfy the following relations:
$$\lim _{\frac{E}{E_{Planck}}\rightarrow 0 } f(\frac{E}{E_{Planck}})=1$$
$$\lim _{\frac{E}{E_{Planck}}\rightarrow 0 } g(\frac{E}{E_{Planck}})=1$$
This condition is needed, as the theory is constrained to reproduce the standard dispersion relation in the longer wavelength limit.

In a recent work \cite{Alexei2015}, a deformed energy-momentum relation has been obtained for the interaction of test particle's
spin with a given gravitational field. The issue of speed of light for spinning particles in an arbitrary electromagnetic field has
been also investigated.

As discussed in details in the work of Mattingly \cite{Mattingly2005}, in general the usual Lorentz invariant dispersion relation
$ E^{2}=m^{2}c^{4}+p^{2} c^{2}$ must be replaced by some function $E^{2}=F(p,m)$. Girelli \cite{Girelli2012} and Amelino-Camelia
\cite{Amelino-Camelia2013} had suggested that, the energy-momentum dispersion relation can be also modified to include its alleged
dependence on the ratio of the particle's energy (E) and the quantum gravity energy ($E_{QG}$).

As discussed by Amelino-Camelia \cite{Amelino-Camelia1998}:

\lq \lq It was realized that extremely precise tests could be made with a sensitivity appropriate to certain order of magnitude
estimates of violations of Lorentz invariance. The sensitivity of the tests arises because there is a universal maximum speed
when Lorentz invariance holds, and even small modifications to the standard dispersion relation relating energy and 3-momentum
give highly magnified observable effects on the propagation of ultra-relativistic particles.\rq \rq\\

With this background the present paper is organized as follows. In Section-\ref{section2}, an expression for Planck Mass or Planck
Energy is derived by equating the Compton wavelength with the gravitational radius of the Kerr rotating body.
In Section-\ref{section3}, a modified dispersion relation is derived. Hence the expression for the variation of photon
propagation speed is derived. In Section-\ref{section4}, it  is found that, the photon propagation speed, depends on the
frequency of the photon, the rotation parameter of the Kerr rotating body and also on the polarization state of the photon.
Finally, some conclusions are made in Section-\ref{section5}.

\section{Planck Mass in Kerr Geometry}\label{section2}

In this section, as a part of the present work, an expression for Planck Mass or Planck Energy is derived by equating the
Compton wavelength with the gravitational radius of Kerr rotating body.

Considering only mass but neglecting the rotation of a body, the geometry of space-time can be described by using the
spherically symmetric Schwarzschild geometry. When mass and rotation both are taken into consideration, the spherical symmetry
is lost and the off-diagonal terms appear in the metric. The
geometry of space-time then can be  described by using the Kerr geometry. The most useful form of the solution of
Kerr metric \cite{kerr1963} is given in terms of t, r, $\theta$ and $\phi$, where t, and r are Boyer-Lindquist coordinates
\cite{Boyer} running from - $\infty$ to + $\infty$, and $\theta$  and $\phi$, are ordinary spherical coordinates in which
$\phi$ is periodic with period of 2 $\pi $ and $\theta$ runs from 0 to $\pi$.

Covariant form of the metric tensor with signature (+,-,-,-) is expressed as:
\begin{equation}\label{30}
ds^{2} = g_{tt}c^{2}dt^{2}+ g_{rr} dr^{2} + g_{\theta\theta} d\theta^{2} +g_{\phi\phi} d\phi^{2} + 2 g_{t \phi} c dt d\phi
\end{equation}
where $g_{ij}$'s are non-zero components of Kerr family.

Non-zero components $g_{tt}$ of Kerr metric is given as follows:
\begin{equation}\label{32}
 (g_{tt})_{Kerr}= (1-\frac{r_{g} r}{r^{2}+a^{2}\ cos^{2}\theta})
\end{equation}
where,
\begin{itemize}
\item $r_g(=2GM/c^2)$ is Schwarzschild radius and
\item M is the mass of the central body,
  \item $ a (= \frac{J}{Mc})$ is rotation parameter of the source,
  \item J is the angular momentum of the central body, which can be also written as $J = I\Omega$,
  \item I is the moment of inertia of the central body, and
  \item $\Omega$ is the angular velocity of the central body.
\end{itemize}

The above equation (\ref{32}) can be written as:
\begin{equation}\label{33}
(g_{tt})_{Kerr}= 1-\frac{r_{g} r}{r^{2}(1+\frac{a^{2}}{r^{2}}\ cos^{2}\theta)}= 1-\frac{r_{g}}{r}(1+\frac{a^{2}}{r^{2}}\ cos^{2}\theta)^{-1}
\end{equation}
Considering $a \ll r$, and using binomial approximation and keeping only the term of order of $\frac{a^{2}}{r^{2}}$,  the
above equation (\ref{33}) becomes:
\begin{equation}\label{34}
 (g_{tt})_{Kerr}= 1-\frac{r_{g}}{r}(1-\frac{a^{2}}{r^{2}}\ cos^{2}\theta)
\end{equation}
By substituting the rotation parameter (a) equals to zero in the above equation (\ref{34}), the component $g_{tt}$ for Kerr
metric reduces to that of Schwarzschild metric,
\begin{equation}\label{35}
 (g_{tt})_{Schwarzschild}= 1-\frac{r_{g}}{r}
\end{equation}
Now comparing the Non-zero component $g_{tt}$ of Kerr metric (as given by equation (\ref{34})) with the component $g_{tt}$ of Schwarzschild metric (as given by equation (\ref{35})), one can conclude that:
\begin{equation}\label{36}
(r_{g})_{Kerr}=r_{g} \ (1-\frac{a^{2}}{r^{2}}\ cos^{2}\theta)
\end{equation}
Here the term $(r_{g})_{Kerr}$ can be identified as the  Kerr gravitational radius for rotating body, analogous to the Schwarzschild
gravitational radius $r_g(=2GM/c^2)$ for static body. The difference between  Kerr metric and Schwarzschild metric is only
due to the effect of rotation. If the rotation parameter (a) goes to zero then from the above equation (\ref{36}), as shown earlier
one can conclude that,
\begin{equation}\label{37}
(r_{g})_{Kerr \ or \ Rotation}=(r_{g})_{Schwrzshild \ or \ Static},  \ if \ a=0
\end{equation}
In other words one can say that,
\begin{equation}\label{38}
(r_{g})_{Kerr \ or \ Rotation}=(r_{g})_{Schwrzshild \ or \ Static} \ (1-\frac{a^{2}}{r^{2}}\ cos^{2}\theta)
\end{equation}
The above equation (\ref{38}) clearly indicates that the effect of rotation is to reduce the Schwarzschild gravitational radius of
the static black hole.

Most of the authors \cite{Yuan2009,Hoyle2000,Crothers2006} had calculated the expression of Planck mass $(M_{Planck})$ by
considering a static body whose Compton wavelength equals to its Schwarzschild gravitational radius. So one can write as;
\begin{equation}\label{x}
  \frac{\hbar}{Mc}=\frac{2GM}{c^{2}}=r_{g}
\end{equation}
From the above equation (\ref{x}), the expression of Planck mass $(M_{Planck})$ can be written as:
\begin{equation}\label{x1}
  (M_{Planck})_{Schwrzshild \ or \ Static} =\sqrt{\frac{\hbar c}{2G}}
\end{equation}

Now one can consider a rotating body whose Compton wavelength equals to its Kerr gravitational radius. Then one can write:
\begin{equation}\label{39}
    \frac{\hbar}{Mc}=\frac{2GM}{c^{2}} \ (1-\frac{a^{2}}{r^{2}}\ cos^{2}\theta)=(r_{g})_{Kerr \ or \ Rotation}
\end{equation}
From the above equation (\ref{39}), the expression for Planck mass $(M_{Planck})$ in Kerr geometry can be re-written as:
\begin{equation}\label{40}
    (M_{Planck})_{Kerr \ or \ Rotation} =\sqrt{\frac{\hbar c}{2G}} \ (1-\frac{a^{2}}{r^{2}}\ cos^{2}\theta)^{-1/2}
\end{equation}

If one considers $a \ll r$, then using binomial approximation and retaining terms only up  to order of $\frac{a^{2}}{r^{2}}$,
the above equation (\ref{40}) becomes:
\begin{equation}\label{41}
    (M_{Planck})_{Kerr \ or \ Rotation} =\sqrt{\frac{\hbar c}{2G}} \ (1+\frac{a^{2}}{2 r^{2}}\ cos^{2}\theta)
\end{equation}

Now the Planck mass $(M_{Planck})_{Kerr \ or \ Rotation}$ as given by the above equation (\ref{41}) is obtained by
considering a rotating body whose Compton wavelength equals to its Kerr gravitational radius.

Now one can write the quantum gravity energy $(E_{QG})_{Kerr \ or \ Rotation}$ as:
 \begin{equation}\label{42}
 (E_{QG})_{Kerr \ or \ Rotation}=(M_{Planck})_{Kerr \ or \ Rotation} \ c^{2}=\sqrt{\frac{\hbar c^{5}}{2G}} \ (1+\frac{a^{2}}{2 r^{2}}\ cos^{2}\theta)
\end{equation}

\section{Modified Dispersion Relation and Variation of Photon Propagation Speed}\label{section3}
In this section using the Planck Mass for Kerr geometry, a modified dispersion relation is derived. Hence an expression for
the variation of photon propagation speed with photon frequency can be derived.

As discussed in the papers \cite{Amelino-Camelia1998,Tawfik}, in case of photon at small energies $E_{Ph} <<E_{QG}$, the
energy momentum dispersion relation can be written as:
 \begin{equation}\label{24}
  c^{2} p^{2}= E_{Ph}^{2}[1+ \xi \frac{E_{Ph}}{E_{QG}}+ o(\frac{E_{Ph}}{E_{QG}})^{2}]
  \end{equation}
Here p is the photon momentum and c is the velocity of light. The quantities $E_{Ph}$ and $E_{QG}$ are photon energy and quantum
gravity energy respectively. The quantity $\xi (= \pm 1)$ is a sign ambiguity that would be fixed in a given dynamical framework.

Using above equation (\ref{24}) and neglecting the second and higher order terms of $(\frac{E_{Ph}}{E_{QG}})$, one can write the
expression for the photon momentum (p) as:
\begin{equation}\label{25}
  p= \pm \frac{E_{Ph}}{c}\sqrt{(1+ \xi \frac{E_{Ph}}{E_{QG}})}
  \end{equation}

Using equation (\ref{24}) and neglecting the second and higher order terms of $(\frac{E_{Ph}}{E_{QG}})$, one can write the
expression for the group velocity (energy-dependent velocities) for photons ($v_{ph}=\frac{\partial}{\partial p}E_{Ph}$) as:
 \begin{equation}\label{26}
   v_{ph}= \frac{\partial}{\partial p}E_{Ph}\approx \frac{p c^{2}}{E_{Ph}(1+\frac{3 \xi E_{Ph}}{2 E_{QG}})}
\end{equation}
Substituting \lq p\rq \ from equation (\ref{25}) in the above equation (\ref{26}), one can write:
\begin{equation}\label{27}
   v_{ph} \approx c   \frac{\sqrt{(1+\xi \frac{E_{Ph}}{E_{QG}})}}{(1+\frac{3 \xi E_{Ph}}{2 E_{QG}})}
\end{equation}
After simplification (using binomial approximation) and neglecting the second and higher order terms of $(\frac{E_{Ph}}{E_{QG}})$,
the above equation (\ref{27}) becomes:
\begin{equation}\label{28}
   v_{ph}\approx c (1-\xi \frac{E_{Ph}}{E_{QG}})
\end{equation}

Now using the equation (\ref{42}), the momentum of photon from equation (\ref{25}) can be re-written as:
\begin{equation}\label{43}
  p= \pm \frac{E_{Ph}}{c}\sqrt{(1+ \xi \frac{E_{ph}}{\sqrt{\frac{\hbar c^{5}}{2G}} \ (1+\frac{a^{2}}{2 r^{2}}\ cos^{2}\theta)})}
  \end{equation}

 Here $E_{ph}(=\hbar \omega)$ is the energy. And $\hbar$  and $\omega$ are  Planck's constant (over 2$\pi$) and angular frequency of photon
 respectively.
 Using equation (\ref{43}), group velocity of photon $(v_{ph})$ given by equation (\ref{28}), can be re-written as:
\begin{equation}\label{44}
   v_{ph}= \frac{\partial}{\partial p}E_{Ph}\approx c (1-\xi \frac{\hbar \omega}{\sqrt{\frac{\hbar c^{5}}{2G}} \ (1+\frac{a^{2}}{2 r^{2}}\ cos^{2}\theta)})
\end{equation}
After simplification the above equation (\ref{44}) becomes:
\begin{equation}\label{45}
   v_{ph}\approx c  (1-\xi \frac{\omega}{c} \ \sqrt{\frac{2\hbar G}{c^{3}}}  \ (\frac{1}{1+\frac{a^{2}}{2 r^{2}}\ cos^{2}\theta}))= c  (1-\xi \ \frac{\omega}{c}    \ \frac{L_{Planck}}{1+\frac{a^{2}}{2 r^{2}}\ cos^{2}\theta})
\end{equation}
Some quantum-gravity theories \cite{Abdo2009,Kostelecky2008,Amelino2009} are consistent with the photon-propagation speed $(v_{ph})$
varying with photon energy. Using high-energy observations from the Fermi Large Area Telescope (LAT) of $\gamma$-ray burst GRB090510,
a model has been tested in which photon speeds are distributed normally around c with a standard deviation proportional to the photon
energy \cite{Vasileiou2015}.

\section{Results and discussion}\label{section4}

From the above equation (\ref{45}) of photon propagation speed $(v_{ph})$, it is clearly seen that, the photon propagation
speed depends on the frequency of the photon, the rotation parameter of the Kerr rotating body and also on the coordinate
$(\theta)$.

The presence of \lq$\theta$\rq \ can be explained in the following way: when light is circularly polarized $cos^{2}\theta=1$ and
when light is unpolarized $cos^{2}\theta=0$. This has justification as,  for the unpolarized light the spin axis of photon is
randomly oriented, the presence of rotation parameter \lq a\rq \ makes no sense. So cos$\theta$ term is zero. Thus from equation
(\ref{45}), two different dispersion relations for circularly polarized light and unpolarized light are obtained. This is a new
property of lorentz invariance, which was not reported by any researcher in past.

Considering the rotation parameter (a) equals to zero, or the latitude $\theta=\frac{\pi}{2}$, (\\ i.e. at the equator),
the above equation (\ref{45}) can be re-written as:
\begin{equation}\label{46}
   v_{ph}\approx c (1-\xi  \ \frac{\omega}{c} \  L_{Planck})
\end{equation}
Considering the latitude, $(\theta=0)$  for North pole, and  $(\theta=\pi)$  for South pole, the above equation (\ref{45}) can be re-written as:
\begin{equation}\label{47}
 v_{ph}\approx  c  (1-\xi \ \frac{\omega}{c} \ \frac{L_{Planck}}{1+\frac{a^{2}}{2 r^{2}}})
\end{equation}

The presence of the Planck length $(L_{Planck}=\sqrt {\frac{2\hbar G}{c^{3}}})$ in the expressions (\ref{45}), (\ref{46}) and
(\ref{47}) of the photon propagation speed $(v_{ph})$, clearly indicates the effect of quantum gravity.

In general the rotation parameter for the Planck sized Kerr rotating body should be quantised as suggested by various authors
\cite{Bakenstein1998,Hod1998,Ropotenko2009}. Notice that, in the present work, the semi-classical theory (for example classical
rotation parameter for the Kerr rotating body) was used, to derive the expression (\ref{45}) for the photon propagation speed.
To study  the variation of photon propagation speed more rigorously, one can use the quantized rotation parameter of planck sized
Kerr rotating body in any future work.

\section{Conclusions}\label{section5}
In the present work, an expression for the Planck Mass or Planck Energy has been derived by considering the equivalence of Compton
wavelength with Kerr gravitational radius of a rotating body. Considering the rotation parameter equals to zero,
one can find that the derived results for Planck mass or Planck energy match very well  with the earlier works available in the
literature (based on the consideration of the equivalence of Compton wavelength with Schwarzschild radius of the static body).
A modified expression for the photon energy-momentum dispersion relation and hence an expression for the variation of the photon
propagation speed with photon frequency has been derived in the present work. It is found that, the photon propagation speed,
depends on the frequency of the photon, the rotation parameter of the Kerr rotating body (which is the photon itself in our case)
and the polarization state of the photon. In general one can say that, we have obtained a new dispersion  relation where
the Lorentz invariance is affected by the polarization state of the photon - an effect which was not previously
reported by any author. The presence of the Planck length in the expression derived for the variation of the photon
propagation speed, clearly indicates the quantum gravity effect.

\section{Acknowledgement}\label{section6}
We wish to thank Dr. Atri Deshmukhya, Department of Physics, Assam University, for very useful suggestions and inspiring discussions. Finally we are thankful to the anonymous referee of this paper, for useful comments.

\end{document}